# ARTICLE

# Chiral oily streaks in a smectic-*A* liquid crystal

Ian R. Nemitz,[a,b,‡] Andrew J. Ferris,[a,‡] Emmanuelle Lacaze[b,a], and Charles Rosenblatt[a,‡‡]

The liquid crystal octylcyanobiphenyl (8CB) was doped with the chiral agent CB15 and spin-coated onto a substrate treated for planar alignment of the director, resulting in a film of thickness several hundred nm in the smectic-*A* phase. In both doped and undoped samples, the competing boundary conditions — planar alignment at the substrate and vertical alignment at the free surface — cause the liquid crystal to break into a series of flattened hemicylinders to satisfy the boundary conditions. When viewed under an optical microscope with crossed polarizers, this structure results in a series of dark and light stripes ("oily streaks") of period ~ 1 μm. In the absence of chiral dopant the stripes run perpendicular to the substrate's easy axis. However, when doped with chiral CB15 at concentrations up to $c$ = 4 wt-%, the stripe orientation rotates by a temperature-dependent angle $\varphi$ with respect to the $c$ = 0 stripe orientation, where $\varphi$ increases monotonically with $c$. $\varphi$ is largest just below the nematic – smectic-*A* transition temperature $T_{NA}$ and decreases with decreasing temperature. As the temperature is lowered, $\varphi$ relaxes to a steady-state orientation close to zero within $\sim 1^{\circ}$ C of $T_{NA}$. We suggest that the rotation phenomenon is a manifestation of the surface electroclinic effect: The rotation is due to the weak smectic order parameter and resulting large director tilt susceptibility with respect to the smectic layer normal near $T_{NA}$, in conjunction with an effective surface electric field due to polar interactions between the liquid crystal and substrate.

## Introduction

Liquid crystals are a fertile source of defect structures[1,2], both in the nematic and the layered smectic phases. Depending on the system's parameters — this includes the composition of the substrate(s), film thickness, applied fields (electric, magnetic, strain), the temperature-dependent elastic moduli, and chirality — the smectic-*A* (Sm-*A*) phase can be deformed into a variety of one- and two-dimensional ordered defect patterns.[3-12] These structures provide insights into the physical behavior of the liquid crystal, such as anchoring properties, phase transitions, and grain boundaries, and are studied for possible applications.

Concomitant with the discovery of liquid crystals in the late 19th century, elongated textures were observed and given the moniker "oily streaks"[13], which is a descriptor of their visual appearance rather than of their chemical composition. The generic oily streak[14,15], which can have complex internal structures, can occur in several liquid crystalline phases, in films of thickness ranging from 100 nm to many micrometers, in closed or open cells, and with different types of boundary conditions at the interfaces. Recent work has focused on a particular category of oily streak that involves extremely thin Sm-*A* films, with one substrate treated for planar alignment of the liquid crystal director $\hat{n}$ and the other interface being open to the air and therefore imposing vertical alignment.[5,8,9,12] This type of oily streak, which is particularly rich in a variety of physical properties, is found in extremely thin (100 < $h$ < 500 nm, where $h$ is the film thickness) smectic films and, as has been demonstrated by optical and x-ray measurements on certain substrates, has the distinctive feature of an ultra-thin transition region, with a planar-oriented director, of smectic layers at the substrate (Fig. 1).[12] In order to match the competing boundary conditions: i) $\hat{n}$ is parallel to the "easy axis" $\hat{e}$ immediately at the substrate over most or all of the alignment layer (except for the very thinnest films; see Fig. 1b for a generic representation), and ii) the smectic layers deform into a series of hemicylinders Fig. 1b, resulting in the texture observed under a polarizing microscope seen in Fig. 1a. It has been shown that for thin films (~120 – 230 nm) of the liquid crystal octylcyanobiphenyl (8CB) on a polyvinyl alcohol (PVA) alignment layer, the director's planar alignment results in an ultra-narrow transition region of smectic layers close to the substrate, where the layers are perpendicular to the substrate plane.[12] Likewise, we can expect the presence of a similar transition region at other types of planar-aligning substrates for extremely thin smectic films that exhibit oily streaks, with layers oriented perpendicular to the substrate covering part, most, or the entirety of the substrate (region B in Fig. 1b). The exact structure of the oily streaks will depend on the choice of liquid crystal, the alignment layer preparation, and most importantly, the thickness of the liquid crystal film.[12] It is this smectic transition region that will be responsible for the

[a.] Department of Physics, Case Western Reserve University, Cleveland, Ohio 44106 USA
[b.] CNRS UMR 7588, Université Pierre et Marie Curie, Institut des NanoSciences de Paris (INSP), 4 place Jussieu, 75005 Paris, France
‡ These authors contributed equally to this work.
‡‡ Author for correspondence. Email: rosenblatt@case.edu.

rotation of the oily streaks when the liquid crystal is doped with a chiral molecule, and which is the subject of this work. Additionally, we note that there is a region (Region A) that could be either a grain boundary or a melted nematic if the director discontinuity is sufficiently large.[12,16] As recent work[5,9,12] has used the simple term "oily streak" to represent this particular variety of oily streak — *viz.*, a very thin Sm-*A* film with an ultra-narrow planar-aligned transition region, hybrid anchoring conditions, and occurring over the entire Sm-*A* temperature range — we will continue that usage herein.

Chirality long has played an important role in soft matter, and liquid crystals in particular. The first recognized liquid crystal, by Reinitzer, involved cholesterol derivatives.[17] Chiral liquid crystals exhibit the so-called "blue phases" that involve multi-dimensional twisted structures[18], are crucial for the ferroelectric properties of liquid crystals[19], and are exploited for their photonic band-gap and lasing properties. [20-23] For sufficiently strong chirality, the Sm-*A* phase can break into regions in which the layer normal rotates about an axis from one region to the next; this is the "twist grain boundary" phase.[24,25] In the Sm-*A* phase, chirality also is required for an electroclinic effect[26], in which the director tilts with respect to the smectic layer normal on application of an electric field in the smectic layer plane.

In this paper we report on oily streaks, which involve rotating smectic layers that can be observed as stripes under crossed polarizers (Fig. 1), and which are composed of the liquid crystal 8CB (Fig. 2a) mixed with small quantities (up to 4 wt-%) of the right-handed chiral dopant CB15 (Fig. 2b). On cooling below the nematic – Sm-*A* phase transition temperature $T_{NA}$, oily streaks without chiral dopant are oriented at an angle $\varphi = 0^o$, *i.e.*, *perpendicular* to the easy axis $\hat{e}$, as is usual.[12] This orientation remains fixed as the temperature is lowered further in the Sm-*A* phase, with the oily streaks present throughout the entire Sm-*A* temperature range. With the addition of chiral dopant, however, $\varphi(T=T_{NA}) \neq 0$, but decreases toward zero as the temperature is reduced. $\varphi$ eventually reaches a steady-state value very close to $0^o$ at a temperature $T_{NA} - T \equiv \Delta T \sim 1^o C$. We ascribe this result to a surface electroclinic effect (ECE), *i.e.*, an ECE in a region very close to the interface, for a chiral Sm-*A* liquid crystal in the bookshelf geometry. The surface ECE for a chiral Sm-*A* phase was first observed by Xue and Clark using a bookshelf liquid crystal geometry.[27] This is an effect in which the director near the substrate tilts by an angle $\theta$ with respect to the smectic layer normal, where $\theta$ is proportional to the polar coupling between the liquid crystal and the substrate — in essence, this polar coupling corresponds to a spontaneous electric polarization localized at, and normal to, the surface. The surface ECE requires sufficiently low symmetry, including the absence of mirror symmetry; in other words, chirality is a required ingredient. More recently Shao, et al observed[28] a giant surface ECE in a 3.5 μm closed smectic cell having a bookshelf geometry, where they found a chiral-induced rotation of dark lines (parallel to the smectic layer normal) below the isotropic – Sm-*A* phase transition temperature. It is important to note that these dark lines do *not* correspond to oily streaks, which are perpendicular to the smectic layer normal in a much thinner open cell.[12] For the cases examined in Refs. 27 and 28, the director's tilt angle $\theta$ was due to the large susceptibility of the director close to the Sm-*A* to smectic-*C* phase transition temperature $T_{AC}$. On the other hand, for our smectic oily streaks we find that the largest oily streak rotation $\varphi$ occurs immediately below the second-order nematic – Sm-*A* transition temperature $T_{NA}$, and falls rapidly toward zero on cooling. We discuss this behavior in the context of the temperature dependence of the smectic order parameter $\psi$ and the director's tilt elastic constant $D$ [$\propto |\psi|^2$] with respect to the smectic layer normal below the second-order nematic – Sm-*A* phase transition temperature.

## Experimental

The liquid crystal 8CB, which has the phase sequence: isotropic — 40.5$^o$ C — nematic — 33.5$^o$C — Sm-*A* — 21.5$^o$C — crystal, was mixed with the right-handed chiral dopant CB15 shown in Fig. 2b. We investigated concentrations $c$ = 0, 1.0, 2.0, 3.4, and 4.0 (± 0.1) wt-%. We also made a 5 wt-% mixture of 8CB with the left-handed chiral dopant ZLI-811 (Fig. 2c) to determine whether the sense of the oily streak rotation would be reversed from that of the right-handed dopant.

A protected aluminum front-surface mirror was cut and then cleaned in detergent, distilled water, acetone, and ethanol. The reflecting surfaces were spin-coated at 1750 rpm with the polyamic acid RN-1175 (Nissan Chemical Industries) and baked according to manufacturer's specifications. The resulting polyimide alignment layer was rubbed with a commercial rubbing cloth (YA-20-R, Yoshikawa Chemical Co.) to obtain an easy axis $\hat{e}$ for planar liquid crystal alignment. Each liquid crystal/chiral dopant composite was mixed as a 0.2 M solution in chloroform and spin-coated at a spinning speed 1750 rpm for 30 s onto the polyimide-coated substrate. Our goal was to obtain liquid crystal films of approximately uniform thickness after the solvent evaporated. Each sample was observed using a polarizing microscope in reflection mode. The use of a mirrored substrate, rather than a transparent microscope slide, greatly enhanced the visibility and definition of the oily streaks. The sample was heated into, and then cooled down from, the isotropic phase, through the nematic phase, and into the Sm-*A* phase just below $T_{NA}$. The approximate film thickness $h \sim 320 \pm 50$ nm was determined by the interference color[12,29]; this corresponds to a slightly thicker film regime than the one investigated in Ref. 12. After $T_{NA}$ was reached, the temperature was reduced in 50 mK steps, allowed to equilibrate, and high resolution (23 megapixels) images (Fig. 1a) were taken at each temperature. Here the polarizer was oriented at 90$^o$ with respect to the analyzer, and at an angle 45$^o$ with respect to the easy axis. Note that the reflected intensity goes to zero uniformly when the polarizer/analyzer is parallel or perpendicular to the oily streaks, in agreement with the picture in Fig. 1b in which the director near the substrate, and therefore the hemicylinders above, rotate in a plane parallel to the surface.

## Results and Discussion

Let us first examine the results qualitatively. Figure 1a shows an image of the oily streaks for the $c$ = 0 wt-% sample, *i.e.*, no chiral dopant, at a temperature $\Delta T$ = 0.38 °C. As is obvious, the streaks are at an angle $\varphi = 0^{\circ}$ with respect to the normal to the easy axis, which corresponds to the horizontal 2 μm length scale bar. It is important to note that the orientation $\varphi$ of the $c$ = 0 oily streaks remained unchanged as the temperature was reduced.

In Fig. 3 we show oily streak images for the $c$ = 4.0 wt-% sample at three temperatures ΔT below $T_{NA}$. Several features can be observed. First and foremost is that the oily streak orientation $\varphi \neq 0$: $\varphi$ is large just below $T_{NA}$ (Fig. 3a) and decreases toward zero with increasing $\Delta T$ (*i.e.,* decreasing temperature) in the Sm-*A* phase. Second, rather than the entire length of the streaks rotating with decreasing temperature, the process occurs by rotation of individual — or a group of — streak *segments*. The ends of the segments are manifested as streak discontinuities, which are especially evident in Fig. 4, where images corresponding to the boxed area in Fig. 3c were collected every 2 s as the sample was cooled at a rate of 1.7 mK $s^{-1}$. As the temperature was lowered the segments broke and melded repeatedly, giving the appearance of an "unzipping" and "zipping" process. Third, we found qualitatively similar results using the left-handed chiral dopant ZLI-811, except that the sense of the oily streak rotation was reversed, *i.e.,* $\varphi < 0$; see Fig. 5. This is an important check to verify that the effect is, indeed, chiral. Fourth, the average periodicity $\langle p \rangle$ of the oily streaks was found to increase with decreasing temperature for all chiral dopant concentrations (including $c$ = 0), eventually reaching a stable periodicity at ΔT $\sim 1^{\circ}$ C. The change in $\langle p \rangle$ appears to occur by means of edge dislocations, similar to those seen in Fig. 4. Although the periodicity is not the focus of this work and will be pursued separately, we believe that the origin of this effect is related to the small degree of smectic order $\psi$ close to $T_{NA,}$ with $\psi$ increasing rapidly as the temperature was reduced.

To extract the rotation angle $\varphi$, we performed a two-dimensional Fourier transform on the images using the software package Image-J®. An example of a real-space image and its Fourier transform are shown in Fig. 6; here $c$ = 3.4 wt.-% and ΔT = 0.35 °C. Because the period of the oily streaks is only slightly larger than optical wavelengths, any sharp features that may exist tend to be washed out, resulting in a near-sinusoidal real space intensity profile in a direction perpendicular to the streaks. As a consequence, all higher harmonics of the fundamental Fourier peak are suppressed; therefore we will focus solely on the fundamental. For each Fourier image we used Image-J's analysis tools to obtain the region at which the Bragg spot is brightest — this generally corresponds to the central part of the fundamental peak — from which we extracted the average rotation angle $\langle \varphi \rangle$, as well as the average wavevector magnitude $\langle q \rangle$. That the Bragg spot is spread out in $q$-space suggests that: i) the film thickness is not completely uniform and there is a range of oily streak spacings, not a single periodicity $p$ associated with each temperature and concentration; ii) there is a range of rotation angles $\varphi$; and iii) some of the spread comes from the segments' ends (the unzipped regions), the edge dislocations, and very wide stripes (Fig. 6a) that tend to appear and become more numerous at lower temperatures. These wide stripes are oriented perpendicular to the easy axis $\hat{e}$ to within an angle of ±0.3°, and their orientation does not change with temperature. Our initial studies of these wide stripes indicate that they have a structure different from the oily streaks, and will be the subject of future research.

Based on the Fourier analysis, Fig. 7 shows the average rotation angle $\langle \varphi \rangle$ *vs.* $\Delta T$ for all concentrations. We also performed a similar measurement using a $c$ = 3 wt.-% mixture on a PVA substrate, having average molecular weight $M_w$ = 85000 – 124000 and rubbed similarly. The results are shown in Fig. 8 for a film of 8CB having a thickness comparable to those shown in Fig. 7.

Several features can be gleaned from Figs. 7 and 8. First, on decreasing the temperature from $T_{NA}$ there is a very rapid decrease of $\langle \varphi \rangle$ immediately below, *i.e.,* within 100 mK, of $T_{NA}$. This behavior occurs not only for the polyimide alignment layer, but also for the PVA alignment layer. The origin of this rapid variation very close to $T_{NA}$ is not clear, although it may be due in part to a combination of temperature gradients across the sample in conjunction with nano-segregation of the chiral dopant that results from a nematic – Sm-*A* phase boundary. Second, on lowering the temperature further by a few hundred millikelvins, the relaxation rate (decrease) of $\langle \varphi \rangle$ with decreasing temperature, *i.e.,* $d\langle \varphi \rangle / dT$, becomes weaker, perhaps even vanishing for the $c$ = 1.0 and 2.0 wt.-%. At still lower temperatures $\langle \varphi \rangle$ relaxes smoothly toward zero. Third, the rotation for the PVA alignment layer is uniformly smaller than for the RN-1175 polyimide alignment layer; this will be discussed later. A more quantitative interpretation of these apparent trends is difficult due to the relative size of the error bars. Nevertheless, the overall trends described above are quite striking.

Xue and Clark demonstrated a smectic surface electroclinic effect, which they ascribed to a polar interaction between the chiral liquid crystal and the substrate that acts like an external electric field localized at the interface.[27] The liquid crystal's molecular orientation at the interface adjusts so that an electric polarization points toward or away from the interface. This, in turn, couples to the director tilt angle $\theta_0$ with respect to the smectic layer normal immediately at the surfaces ($z$ = 0), which relaxes elastically (by an azimuthal twist) into the bulk. (See Fig. 9) The average value of $\theta_0$ at the surface would vanish if the liquid crystal were achiral, but $\theta_0$ is nonzero with a particular sense of rotation (clockwise or counterclockwise) that depends on the handedness of the liquid crystal (*cf.* Figs. 3 and 5). Xue and Clark's free energy calculation[27] for $\theta(z)$ was based on that for the bulk ECE[26], but with significant modifications to account for its localization near the surface. First, there is a characteristic correlation length $\xi = (K_{22}/D)^{1/2}$ over which the director tilt with respect to the smectic layer normal, *viz.* $\theta(z) = \theta_0 \exp(-z/\xi)$, relaxes from $\theta_0$ at the surface to $\theta = 0$ in the bulk (Fig. 9). Here $K_{22}$ is the twist elastic constant and $D^{-1}$ is an effective susceptibility for the tilt of the director relative to the layer normal, where $D$ vanishes on approaching $T_{AC}$. Second, the applied electric field for the bulk ECE[26] corresponds to a localized field due to the polarization discontinuity at the interface[27] for the surface ECE. For sufficiently small $\xi$, Xue and Clark found that the induced director tilt at the surface $\theta_0$ is proportional to $\xi/K_{22}$, where the proportionality coefficient depends on the generalized polarization susceptibility, the strength of the polar interaction at the interface, and coupling between tilt angle and polarization. Because we are doping our configurationally achiral liquid crystal with a chiral additive, the coupling in our experiment is also related to the dopant concentration $c$, increasing with increasing $c$.

Turning to our experiments, which involve a complex layer structure (Fig. 1b) rather than the uniform Sm-*A* phase studied in Ref. 24, the following physical picture emerges: The smectic layers over most of the surface (item B in Fig. 1b) are perpendicular, or nearly so, to the alignment layer. Because of the presence of an easy axis $\hat{e}$, the director $\hat{n}$ at the substrate is approximately parallel to $\hat{e}$. We remark that this is different from the case of Ref. 27, where the director orientation was established by shearing the cell. Then as with Ref. 27, the polar interaction between the liquid crystal and the alignment layer, *i.e.*, the effective surface field, induces a rotation of the director $\theta_0$ with respect to the smectic layer normal at the surface (Fig. 9). This is the surface ECE, which can be large just below $T_{NA}$ because the elastic constant $D$ for director tilt is proportional to $|\psi|^2$ (Refs. 30 and 31). But owing to the easy axis, it is the director *at the surface* that remains fixed and it is the smectic layers that rotate; again, see Fig. 9. This is the mechanism that drives the rotation of the oily streaks, which must be perpendicular to the smectic layer normal.

As one moves a small distance away from the surface and into the bulk where the effective surface field vanishes, the director orientation relaxes over a length scale $\xi$, so that $\hat{n}$ becomes perpendicular to the already-rotated smectic layers. Shao, et al noted[28] that this variation of $\theta$ with $z$ would entail a small amount of smectic layer curvature near the surface. In addition, they noted that there must be a periodic smectic layer dislocation at the surface due to the difference in smectic layer spacing $d = d_0 cos\theta(z)$ between surface and bulk, where $d_0$ is the smectic layer spacing corresponding to the director being normal to the layer. These effects are likely to be small given our relatively small values of $\theta$.

Returning to the chiral oily streaks, on reducing the temperature below $T_{NA}$ and going more deeply into the Sm-*A* phase, the smectic order parameter $\psi$ increases. This has the effect of increasing the tilt elastic coefficient $D$ and thereby decreasing $\theta_0$ at the interface. Because the director remains parallel to the easy axis $\hat{e}$, the smectic sublayers (item B in Fig. 1b) and, in consequence, the oily streaks must rotate so that the oily streaks approach an orientation perpendicular to $\hat{e}$ at low temperatures. Since rotation of the long oily streaks *en masse* would involve significant defect movement at points far from the pivot, the oily streaks break into smaller segments that can rotate more easily, unzipping and zipping at the segment ends.

Xue and Clark's experiments[27] were performed near $T_{AC}$, and therefore the relevant correlation length was based on the tilt susceptibility $D^{-1}$ diverging at $T_{AC}$. Our experiments take place in a very different region, just below the nematic – Sm-*A* transition temperature, where $D \propto |\psi|^2$ and $\psi$ grows with decreasing temperature. Litster, *et al* used light scattering to obtain the quantity $D/K^0_{33}$ *vs.* temperature in the Sm-*A* phase for the liquid crystal octyloxycyanobiphenyl (8OCB)[32], where $K^0_{33}$ is the non-divergent part of the bend elastic constant. They found that $D/K^0_{33}$ evolves from approximately 3 x $10^{11}$ $cm^2$ at $\Delta T \sim 30$ mK to approximately $10^{12}$ $cm^2$ at $\Delta T \sim 300$ mK to approximately 3 x $10^{12}$ $cm^2$ at $\Delta T \sim 3^oC$. Accounting for the smaller value — by at least a factor of 3 (Ref. 31) of the twist elastic constant $K_{22}$, we would expect $\xi$ to be of order 100 nm at $\Delta T \sim 30$ mK and of order 35 nm at $\Delta T \sim 1^oC$. To be sure, 8CB and 8OCB are not identical molecules and their critical behaviors differ.[32] Moreover, these lengths already are larger than the typical 20 nm thickness of the smectic transition region at the base of the hemicylinders obtained on PVA substrates[12] (region B in Fig. 1b). In consequence, the decaying exponential form $\theta(z) = \theta_0 \exp(-z/\xi)$ is not totally appropriate for the oily streak situation. Nevertheless, the qualitative features that we observe are consistent with the surface electroclinic effect: in particular, a rotation of the oily streaks just below $T_{NA}$ that is larger with higher chiral dopant concentration $c$, which decreases as the temperature is reduced in the Sm-*A* phase.

Let us briefly address the magnitudes observed for the average oily streak rotation angle $\langle\varphi\rangle$. First, values of $\langle\varphi\rangle$ measured for the PVA alignment layer are smaller than the corresponding values for the polyimide, although the general behavior with temperature is qualitatively similar. This may be expected, as the polar interactions that drive the ECE are likely to be different for the two alignment materials; it also is possible that the transition regions close to the substrates have different structures. Second, values of $\langle\varphi\rangle$ observed by Xue and Clark[27], which were due to the proximity of the Sm-*C* phase, were relatively small. The rotation angles observed by Shao, et al[28], also with a nearby Sm-*C* phase, were considerably larger owing to their use of the compound W415, which has an enormous surface ECE. Their pure chiral liquid crystal did not possess a nematic phase and therefore there was no nematic – Sm-*A* transition at which the smectic order parameter $\psi$ would vanish. This is why no decrease of $\langle\varphi\rangle$ was observed as the temperature was reduced in the Sm-*A* phase in Ref. 28, in contrast with chiral oily streaks discussed herein. The relatively large rotation angles $\langle\varphi\rangle$ that we observed just below $T_{NA}$, especially for the polyimide alignment layer (Fig. 7), occur even though our chiral dopant concentration is only a few wt.-%. We speculate that our large rotations may be due to partial segregation of the chiral dopant near the surface. This would facilitate a larger tilt of the director $\theta_0$ relative to the layer normal than would occur if the chiral dopant concentration were uniform throughout the liquid crystal film. This conjecture requires further study.

## Conclusions

Chirality plays an important role in many fields. Here we have shown that a chiral dopant causes a rotation by angle $\varphi$ of smectic oily streaks relative to a direction perpendicular to the easy axis. The rotation angle is largest just below the nematic – Sm-*A* transition temperature, and relaxes toward zero with decreasing temperature in the Sm-*A* phase. The results are consistent with a surface smectic electroclinic effect, albeit due to the large director tilt susceptibility close to $T_{NA}$ (rather than the Sm-*A* – smectic-*C* transition temperature), where the Sm-*A* order parameter is small. The controlled tilt of the oily streaks demonstrated in this work also can be viewed as a potential new tool for the control of topological defect orientation, especially in light of the large number of topological defects known to be present in the oily streak structure above the thin transition region.[12] It could even allow for rotation of nanoparticle assemblies trapped in these defects.[33,34]

## Acknowledgements

IRN and CR were supported by the National Science Foundation's Condensed Matter Physics program under grant DMR-1505389. AJF was supported by a grant from the US-Israel Binational Science Foundation. EL was supported by Centre National de la Recherche Scientifique and Agence Nationale de la Recherche under program « blanc 2013 » (ANR grant 2013-BSV1-0026-03). IRN's, CR's, and EL's travel between the US and France was supported by the Partner University Fund, administered by the French consulate in New York.

## Notes and references

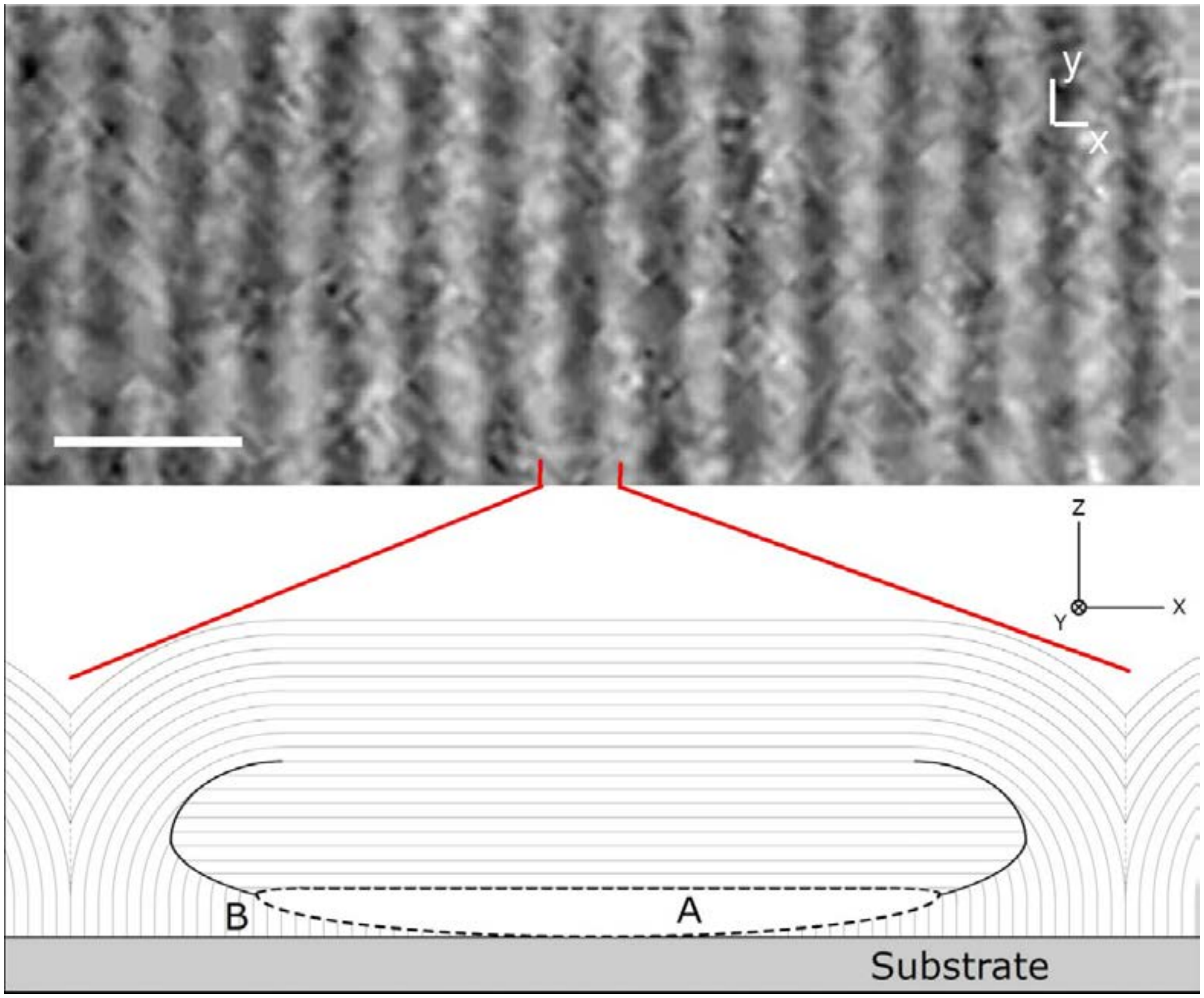

Fig. 1 a) Photomicrograph of oily streaks looking downward along the *z*-axis for undoped liquid crystal at temperature $T_{NA}$ - $T$ = 0.38 °C. Polarizer and analyzer are crossed and are at an angle of 45° with respect to the easy axis orientation. Horizontal bar corresponds to 2 μm and is parallel to the easy axis . b) Generic schematic diagram of an end-on view of the liquid crystal structure that forms oily streaks. The red lines indicate the portion of (a) shown schematically in (b). The precise structure depends on the elastic constants, $T_{NA} - T$, and the film thickness $h$. For this work the most important feature is the orientation of the smectic layers (Region B) at the substrate.

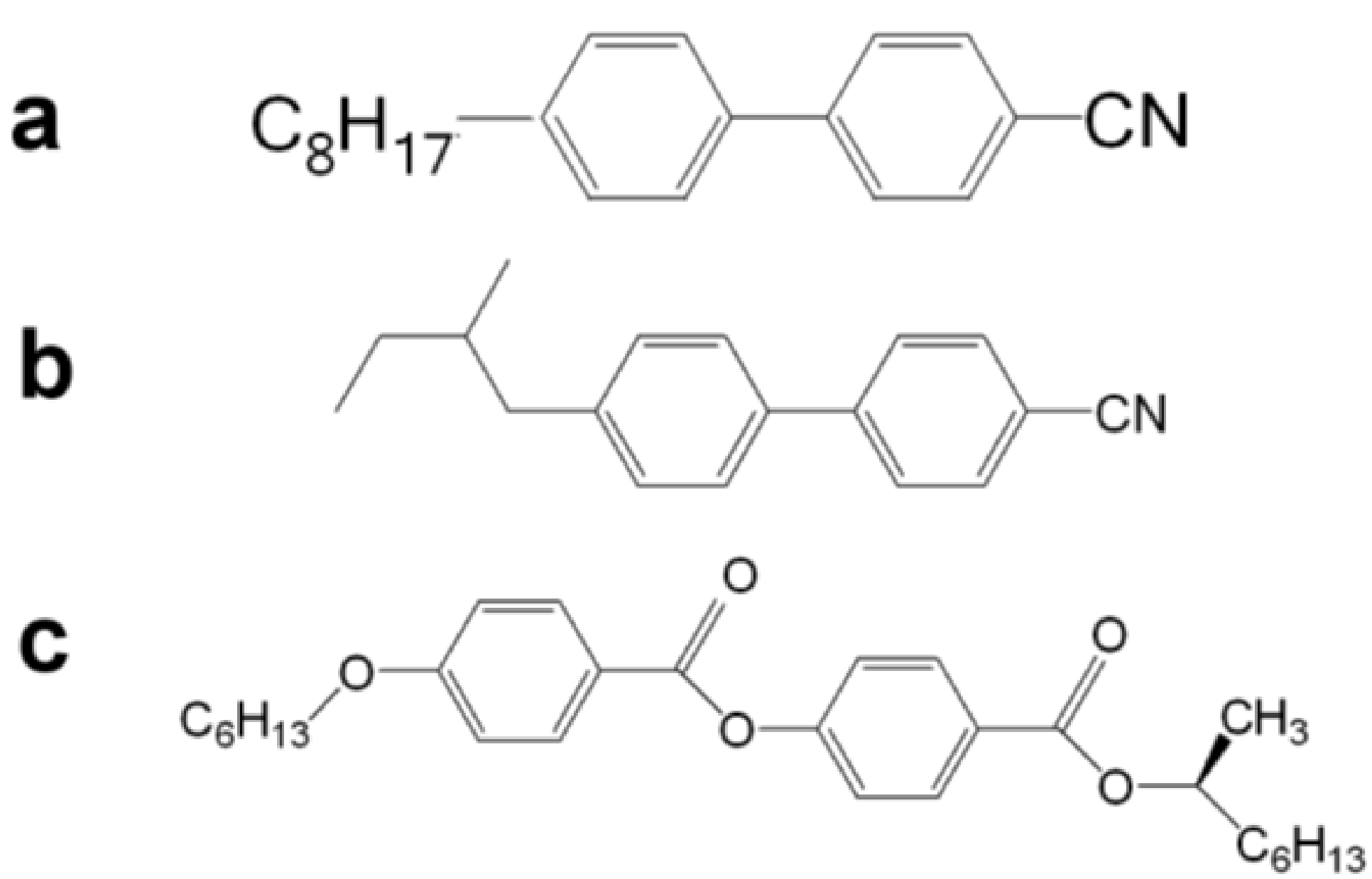

**Fig. 2** Molecules used in this work: a) the liquid crystal 8CB, b) the right-handed chiral dopant CB15, and c) the left-handed chiral dopant ZLI-811.

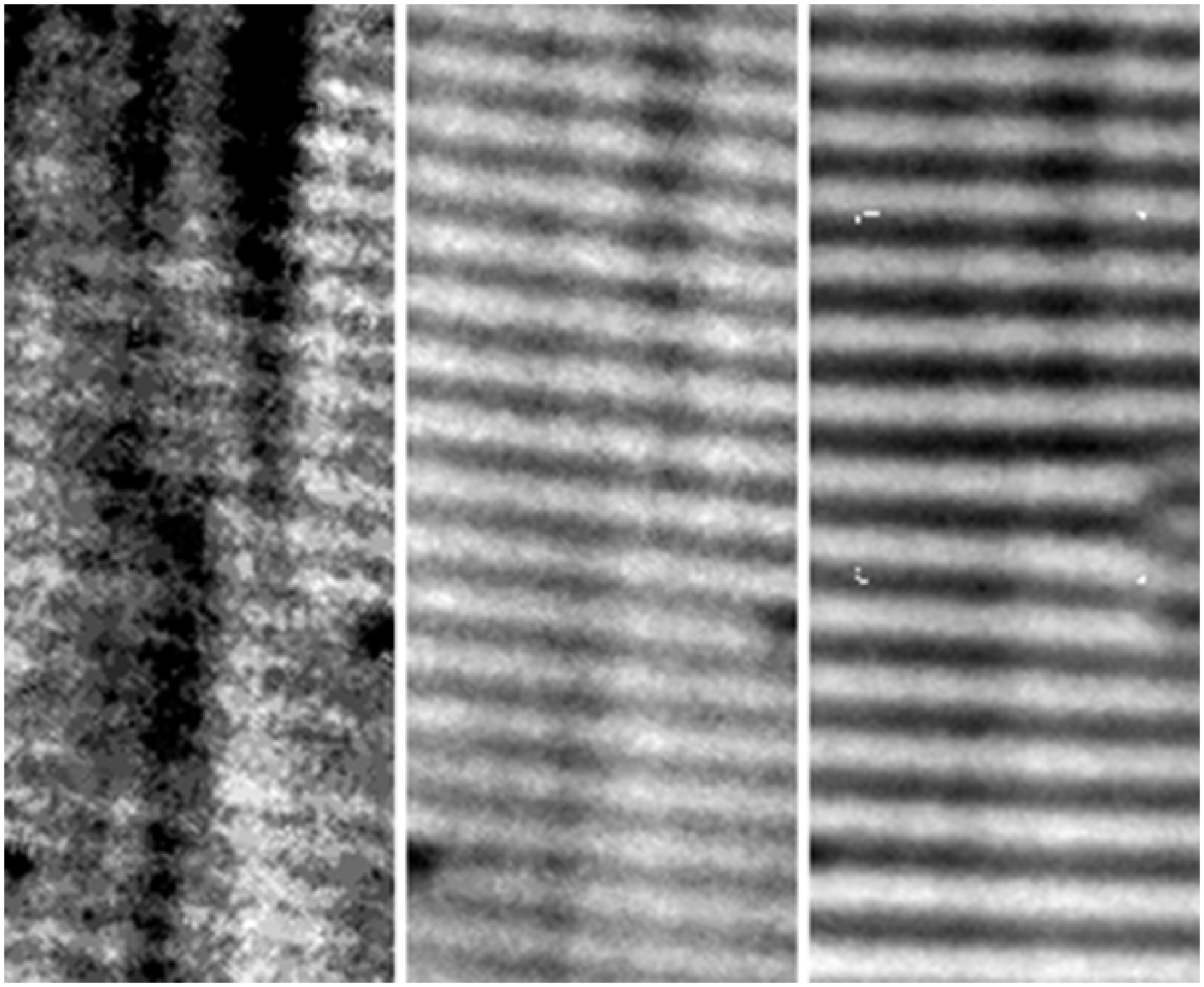

Fig. 3  Photomicrographs of oily streaks for $c$ = 4 wt-% at temperatures $\Delta T$ = 0.2 (a), 0.4 (b), and 1.3 (c) $^{o}$C. Note that the first image (a) has been contrast-enhanced in order to make visible the rotated oily streaks, as the optical contrast close to $T_{NA}$ is particularly weak. The bar represents 2 $\mu$m and is parallel to the easy axis. The dashed rectangle corresponds to the area displayed in Fig. 4.

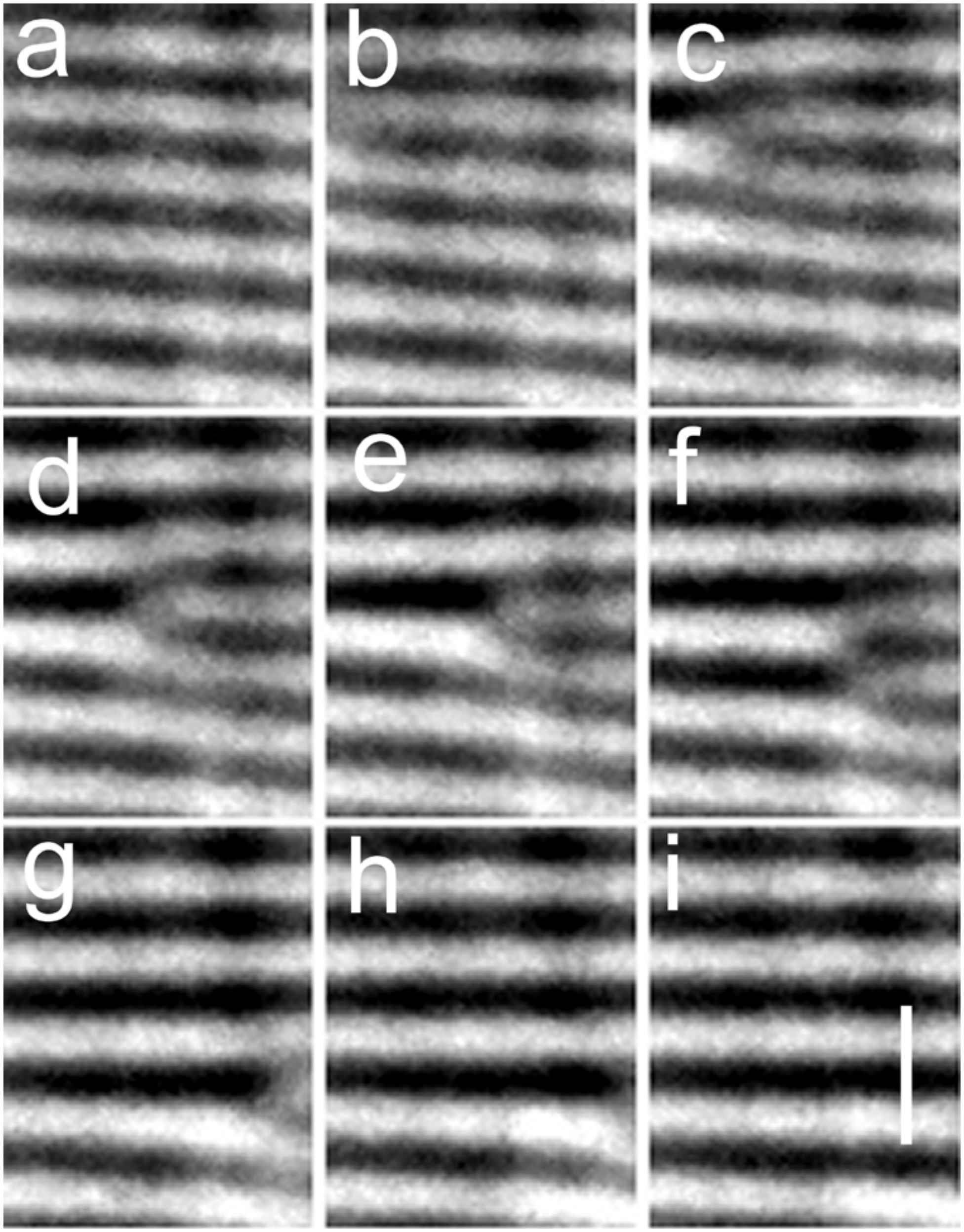

Fig. 4 A series of photomicrographs of oily streaks for *c* = 4 wt-% in the rectangular region of Fig. 3c. Images were taken at 2 s intervals (each with a 1 s exposure time) as the temperature was reduced at a rate of approximately 1.7 mK $s^{-1}$ in the neighborhood of $\Delta T = 1.3^{o}$ C. Notice that the discontinuities in the streaks move downward and to the right as the sample is cooled. Bar in image *i* represents 2 μm and is parallel to the easy axis

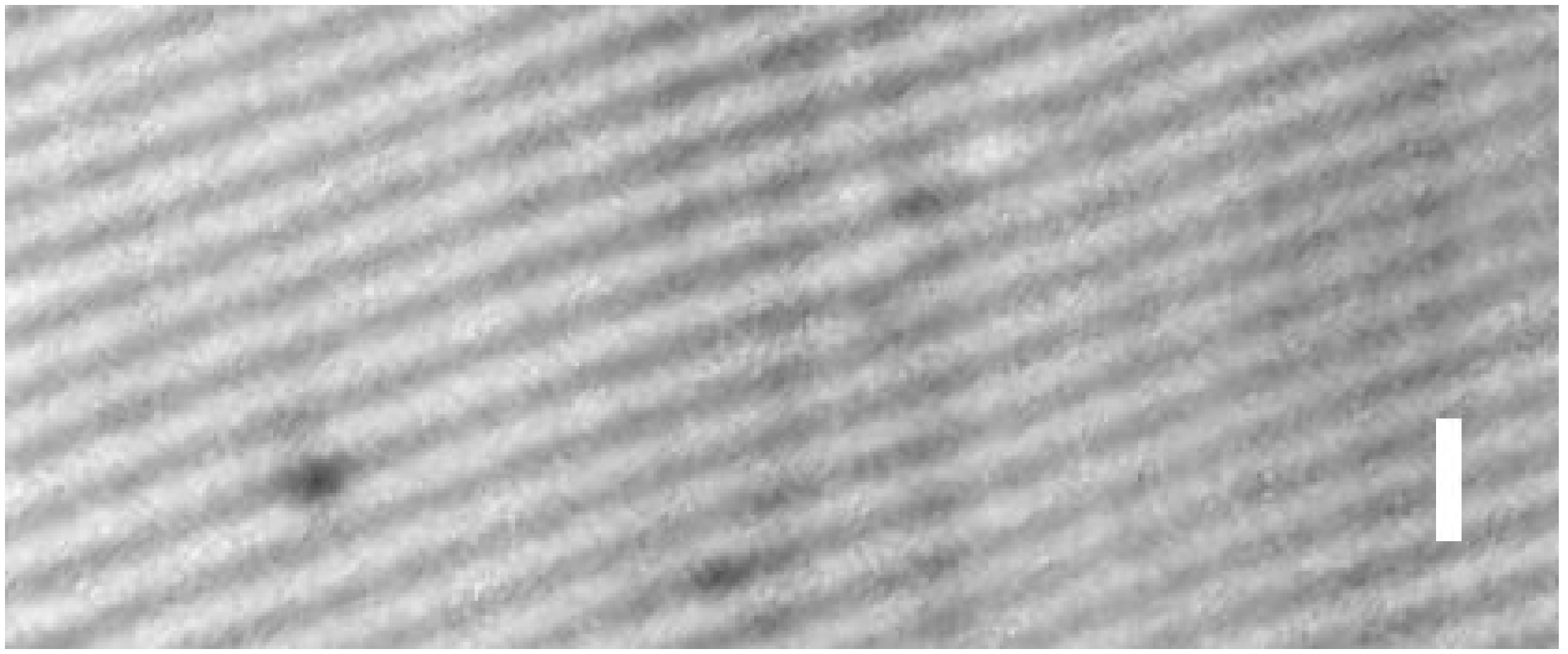

Fig. 5 Photomicrograph of oily streaks having concentration *c* = 5 wt-% of the *left-handed* chiral dopant ZLI-811. Notice that the rotation of the oily streaks has a sense opposite that of the right-handed dopant CB15 in Fig. 4. Bar corresponds to 2 μm and is parallel to the easy axis.

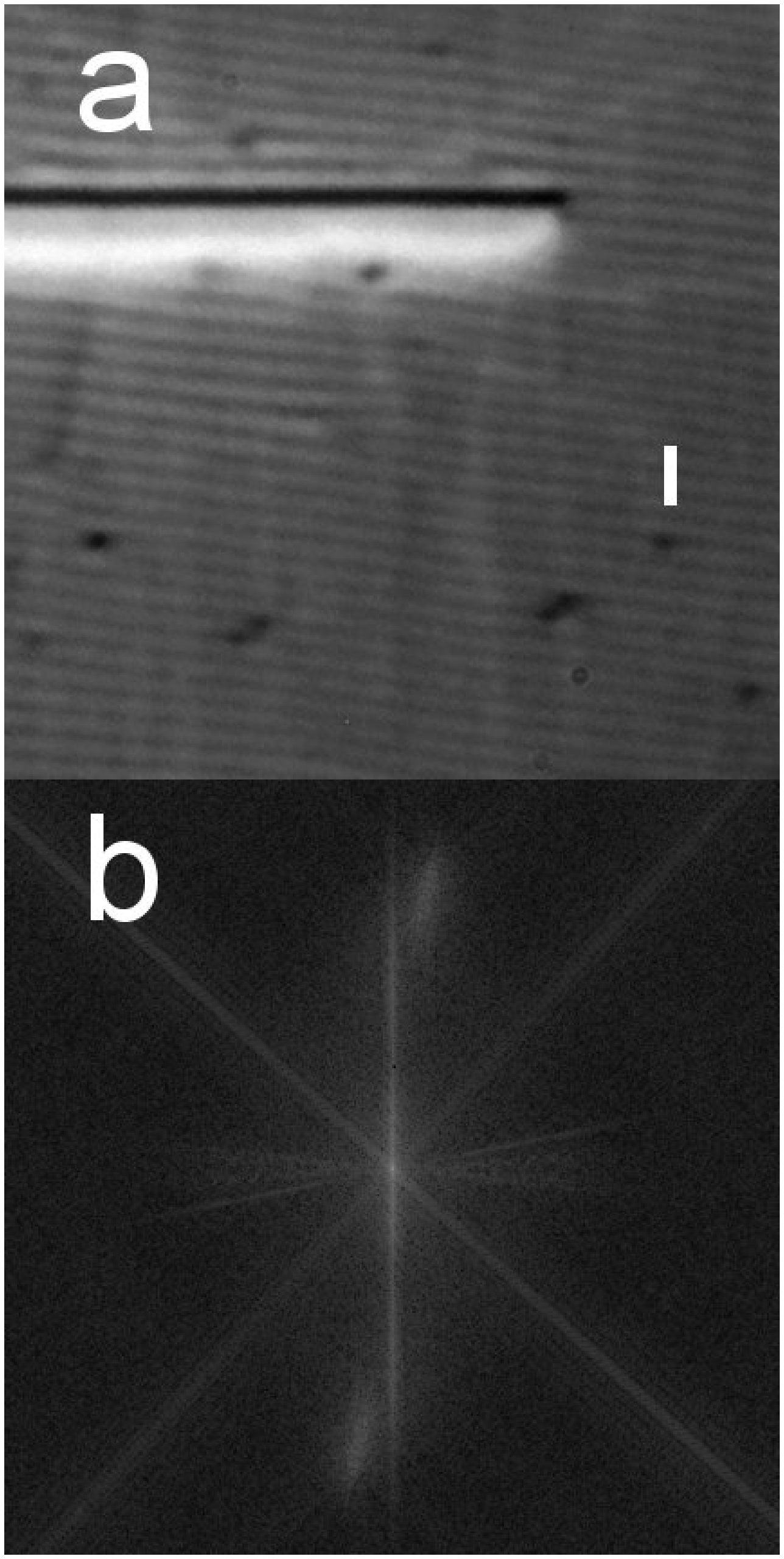

Fig. 6 (a) Real space image of the c = 3.4 wt.-% sample at $\Delta$T = 0.35 $^{o}$C. Bar represents 2 $\mu$m and is parallel to the easy axis. The wide stripe is discussed in the text. (b) 2D Fourier transform of (a). The lines at ±45$^{o}$ are image compression artifacts, the approximately horizontal lines are artifacts of the alignment layer rubbing to create the easy axis, and the vertical line artifact is due to the wide stripe. The relevant features, which are easy to extract, are the elongated smudges in the first and third quadrants, from which we determined the average oily streak rotation angle $\langle\varphi\rangle$ and average periodicity $\langle p\rangle$

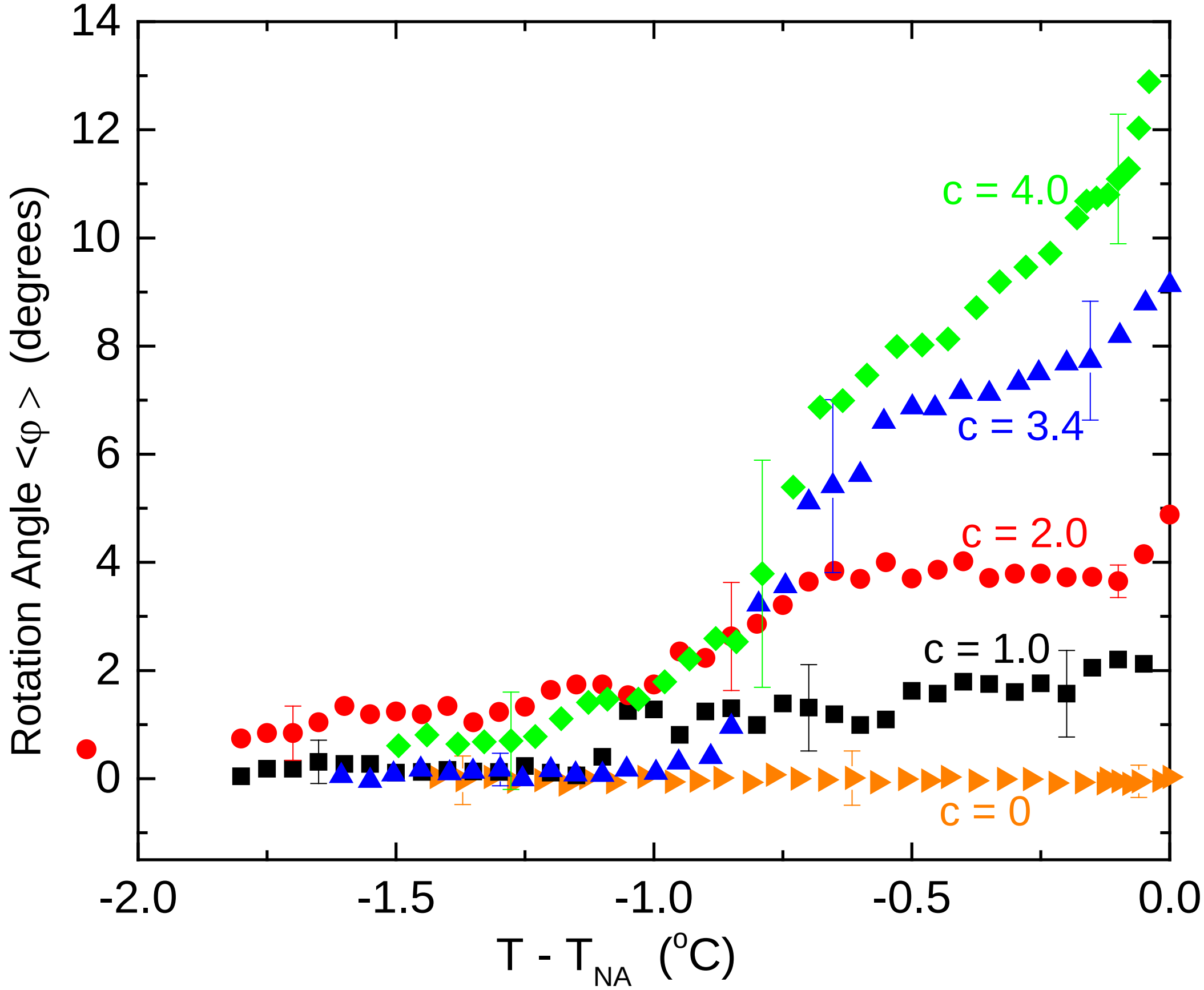

Fig. 7 The average rotation angle (relative to the wide stripes (Fig. 6a)) *vs.* $T - T_{NA}$ for all concentrations decreases as temperature is reduced, approaching a limiting value that corresponds to the wide stripes (Fig. 6a) that appear at lower temperatures and remain fixed in orientation as the temperature is further reduced. Typical error bars are shown.

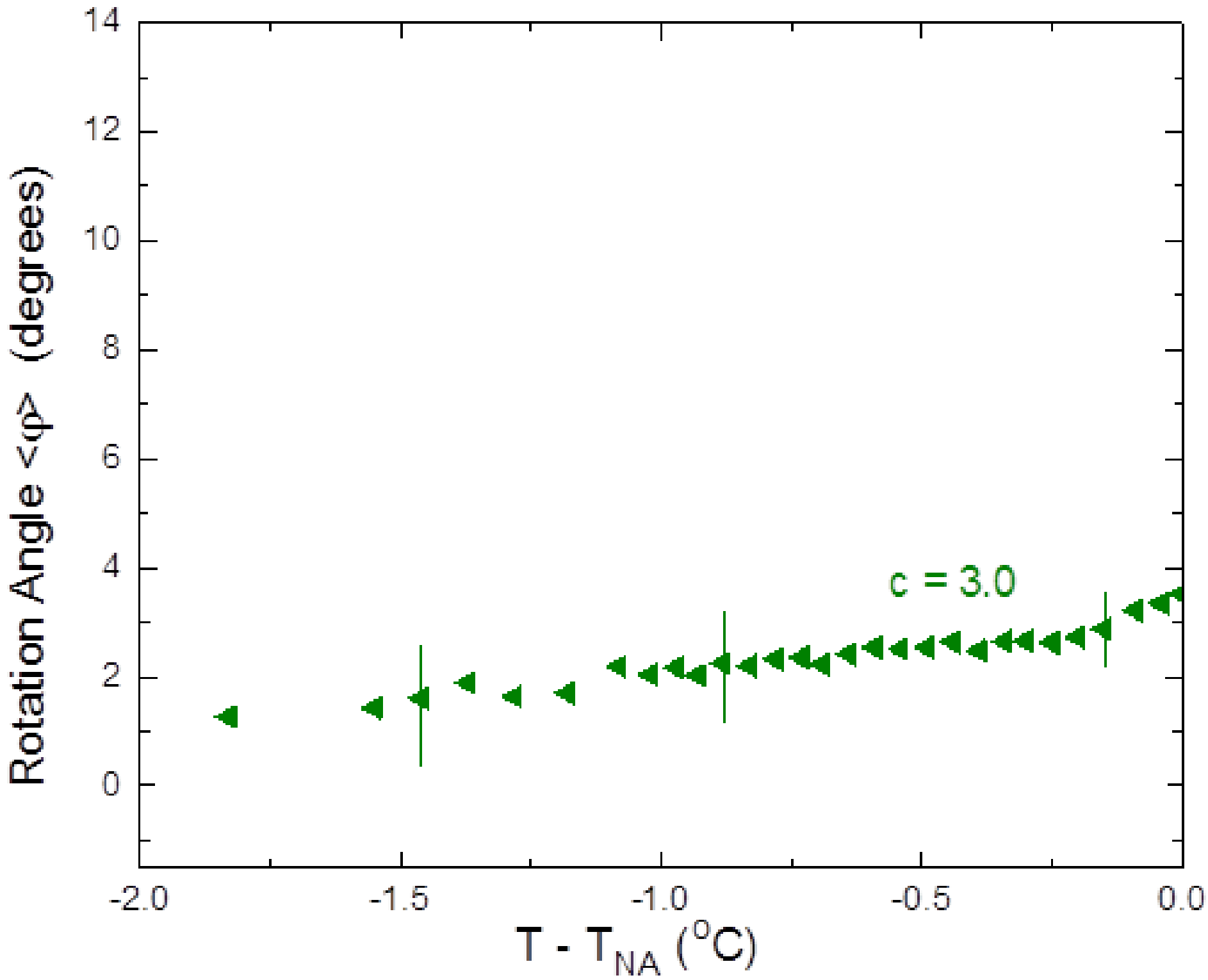

Fig. 8 Same as Fig. 7, but for a PVA substrate. Concentration c = 3 wt.-%

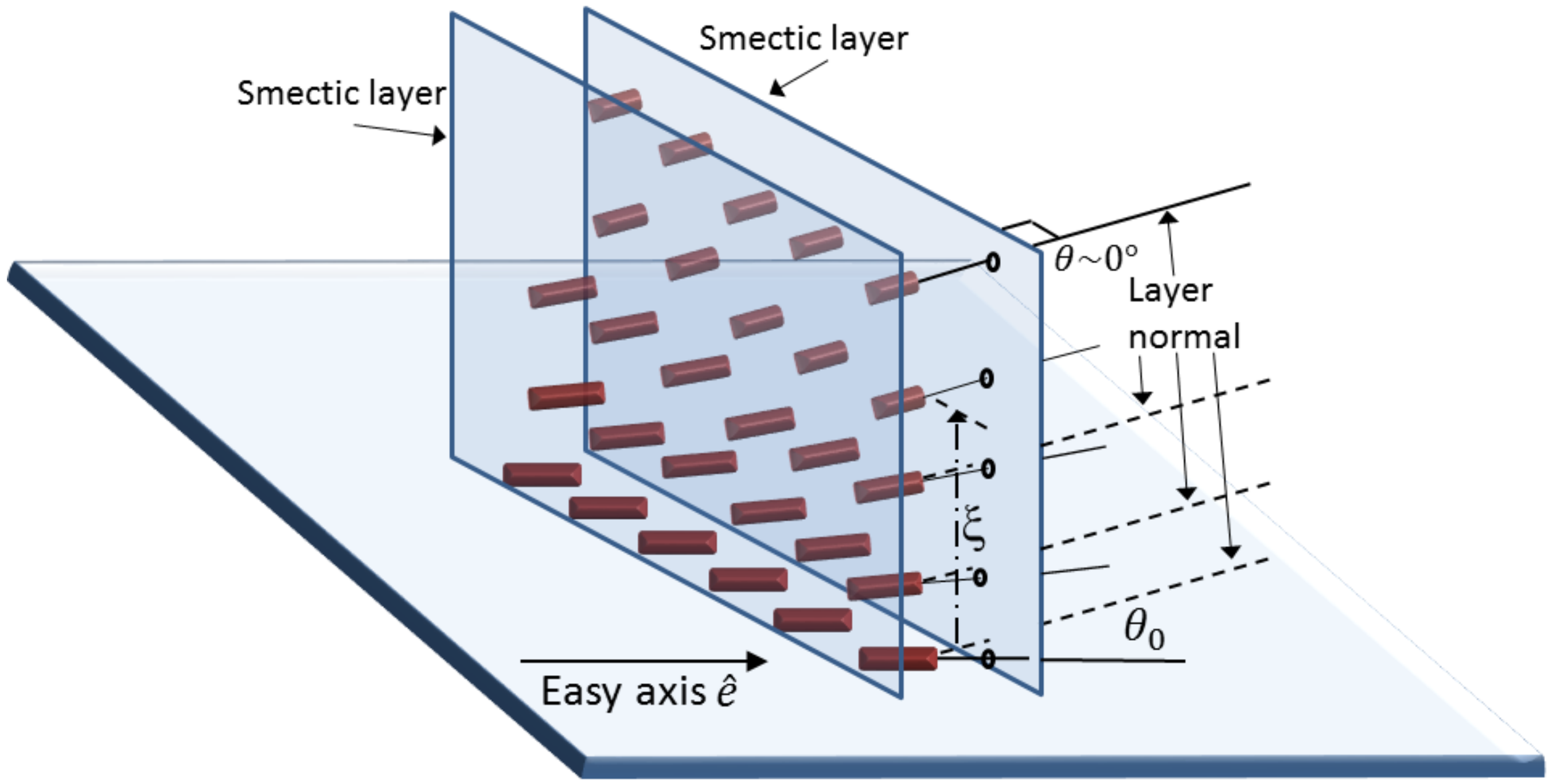

Fig. 9 Schematic diagram of director and smectic sublayers (item B in Fig. 1b) close to the alignment layer. Director is parallel to the easy axis at the substrate ($z$ = 0), where the smectic layers are rotated by an angle $\theta_0$ with respect to the director due to the ECE. The director orientation relaxes to be parallel to the smectic layer normal over a distance $z = \xi$ from the substrate. The distance of the molecules above the substrate is represented by the darkness (color saturation), with the darkest (deepest red) being closest to the substrate and the lightest (least saturated red) being far above the substrate.